# Decentralized Dynamic State Estimation in Microgrids

Bang L. H. Nguyen
ECE Department
Clarkson University
Potsdam, NY, USA
bangnguyen@ieee.org

Tuyen V. Vu
ECE Department
Clarkson University
Potsdam, NY, USA
tvu@clarkson.edu

Tuan A. Ngo
Electric Power Engineers, Inc.
Texas, Austin
ngo.tuan.1985@gmail.com

*Abstract*—This paper proposes a decentralized dynamic state estimation scheme for microgrids. The approach employs the voltage and current measurements in the *dq0* reference frame through phasor synchronization to be able to exclude orthogonal functions from their relationship formulas. Based on that premise, we utilize a Kalman filter to dynamically estimate states of microgrids. The decoupling of measurement values to state and input vectors reduces the computational complexity. The Kalman filter considers the process noise covariances, which are modified with respect to the covariance of measured input values. Theoretical analysis and simulation results are provided for validation.

*Index Terms*— Dynamic state estimation, Kalman filter, microgrids, ship power systems.

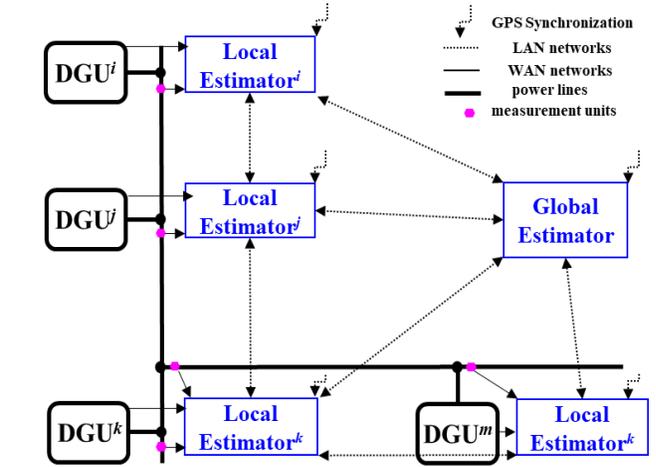

Figure 1. Proposed decentralized dynamic state estimation scheme.

## I. INTRODUCTION

U.S. Navy ship power systems can be seen as microgrids. The knowledge of microgrid states is necessary for controls, protections, contingencies analysis, and load shedding [1]. Navy ships have many challenges as they are designed to be resilient to damages. Damages can be the results of cyber/physical attacks from the enemies. To be resilient to damages, situational awareness of the system via system state estimation is essential for optimal management. Unlike the transmission lines, microgrids behavior changes swiftly that the traditional static estimators may not be able to capture the high-fidelity states. In addition, the integration of distributed energy resources, storage devices are changing the distribution system and making it more and more dynamic. Hence, the real-time monitoring system observing the system dynamics instead of the static ones should be developed to improve situational awareness accuracy [2].

In low voltage distribution grids, the power measurements are usually substituted by ampere measurements for better inspection of overloads and protection from short-circuiting. In addition, the R/X ratios are much higher than those of transmission systems, and the line susceptances can be negligible unless there is a capacitor bank. These characteristics lead to the problem of power decoupling and the complexity in the relation equations of currents and voltages in the polar form [3]. To solve this problem, [4]-[6] introduced an estimated algorithm based on rectangular form or in three-phase. However, the orthogonal functions that caused nonlinearity still presented in the formulas.

Thanks to the development of intelligence electronics devices (IED) and micro-phasor measurement units (μPMU) for distribution grids, the monitoring points associated with protection can be increased with less expensive cost [7]. Besides, distributed generation units (DGU) for their own control requirement also provide the information of the power flows, terminal voltages and currents, which are measured with high accuracy [8]. As a result, the observability in microgrids can be improved.

Generally, the centralized dynamic state estimation requires a high bandwidth communication channel due to the large data set of measurements transferring to the control center. This situation may inhibit the sampling rates of modern sensing devices not to overwhelm the network. The distributed estimation schemes can be an effective solution to this problem. Multi-areas with overlapping subsystems [9]-[11] and decomposition of measurements [12] are the popular approaches. In [13], the matrix slitting techniques were implement based on the inherent sparse structure of power systems. [14] proposed a dynamic state algorithm by sharing the neighboring gains among nearby local estimators to achieve state consensus. In [15], the small signal model of islanded microgrid was employed for estimation the states at each distributed generation without communication.

This paper proposed a decentralized dynamic state estimation scheme shown in Fig. 1 based on voltage and current measurements in the same rotating reference *dq0*-frame through synchronization. The key characteristics of this scheme are listed as follows:

1) The measurement values of voltages and currents are obtained from µPMU with high precision [8].
2) The orthogonal functions (e.g. sin, cos) are excluded from the relationship of voltages and currents provided that they are transferred into the *dq0*-reference frame with the same rotation speed through the GPS synchronization.
3) The local estimators are independent of each other and the global estimator. Hence, their computation processes can be rapidly performed in parallel and even at different sampling rates.
4) The local estimators are considered to install in conjunction with the DGU and critical loads with a high sampling rate for the quick response of control and protection in their own area. The local estimators treat the input signals as disturbances. However, they can receive the neighbor states from the directly connected local estimators for bad data detection.
5) All local states and other information are processed by the global estimator to provide the real-time big picture of the entire microgrid.

The remaining contents of this paper are organized as follows. In Section II, the dynamic model of microgrids is described. Section III illustrates the decentralized dynamic state estimations. In Section IV, the simulation results are provided and discussed. Finally, Section V concludes the paper.

## II. MICROGRIDS DYNAMIC MODEL

For the sake of simplicity, a microgrid system with three DGUs and three buses as shown in Fig. 2 will be taken into account for modeling and analysis. The derived models will be used for the local estimators for the state estimation within the DGU areas, and the global estimator for the states of the distribution lines. Note that, the currents flowing out of the bus associated with DGU bus are treated as disturbances in both DGUs and the microgrid. The dynamic state-space models of microgrid components are given in the following subsections. Note that *i*, *j* denote the bus numbers; $(\cdot)_{abc}$ denotes separately the reference frame $((\cdot)_a, (\cdot)_b, (\cdot)_c)$; $(\cdot)_{dq}$ denotes as $((\cdot)_d + j(\cdot)_q)$.

### A. Distributed Generation Buses

The dynamical equations in *abc*-frame of DGU circuits are

$$\frac{d}{dt}i_{ti,abc} = -\frac{R_{ti}}{L_{ti}}i_{ti,abc} + \frac{1}{L_{ti}}\left(v_{ti,abc} - v_{i,abc}\right), \quad (1)$$

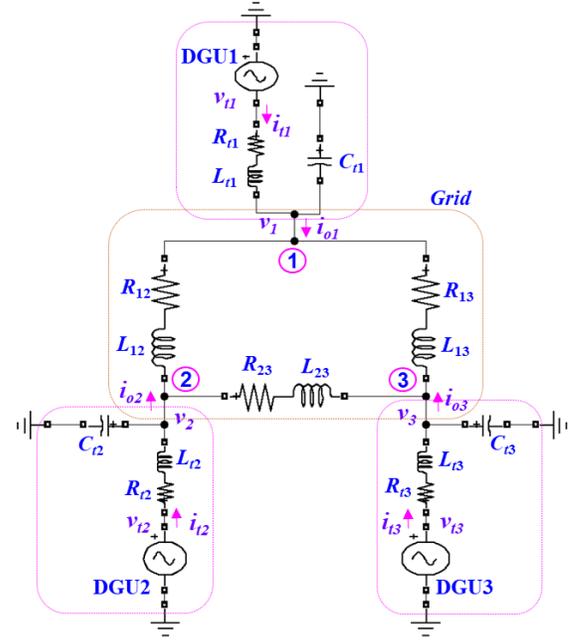

Figure 2. Electrical diagram of a three-DGU three-bus microgrid system.

$$\frac{d}{dt}v_{i,abc} = \frac{1}{C_{ti}}\left(i_{ti,abc} - i_{oi,abc}\right). \quad (2)$$

By applying the *abc/dq0* Park transformation in the same rotating frame of $\omega t$, the below equations are achieved:

$$\frac{d}{dt}i_{ti,dq} + j\omega i_{ti,dq} = -\frac{R_{ti}}{L_{ti}}i_{ti,dq} + \frac{1}{L_{ti,dq}}\left(v_{ti,dq} - v_{i,dq}\right), \quad (3)$$

$$\frac{d}{dt}v_{i,dq} + j\omega v_{i,dq} = \frac{1}{C_t^i}\left(i_{ti,dq} - i_{oi,dq}\right). \quad (4)$$

In the steady-state condition, the derivative terms are zero, and the following relationships are satisfied:

$$v_{ti,dq} - v_{i,dq} = R_{ti}i_{ti,dq} + j\omega L_{ti}i_{ti,dq}, \quad (5)$$

$$i_{ti,dq} - i_{oi,dq} = j\omega C_{ti}v_{i,dq}. \quad (6)$$

From the above equations, the continuous-time state space model of DGU buses can be determined as:

$$\begin{cases} \dot{x}_{Gi}(t) = A_{Gi}x_{Gi}(t) + B_{Gi}u_{Gi}(t) + w_{Gi}(t) \\ y_{Gi}(t) = C_{Gi}x_{Gi}(t) + v_{Gi}(t) \end{cases}, \quad (7)$$

$$A_{Gi} = \begin{bmatrix} 0 & \omega & 1/C_{ti} & 0 \\ -\omega & 0 & 0 & 1/C_{ti} \\ -1/L_{ti} & 0 & -R_{ti}/L_{ti} & \omega \\ 0 & -1/L_{ti} & -\omega & -R_{ti}/L_{ti} \end{bmatrix}, \quad (8)$$

$$B_{Gi} = \begin{bmatrix} 0 & 0 & -1/C_{ti} & 0 \\ 0 & 0 & 0 & -1/C_{ti} \\ 1/L_{ti} & 0 & 0 & 0 \\ 0 & 1/L_{ti} & 0 & 0 \end{bmatrix}. \quad (9)$$

The state and input vectors are respectively as follows:

$$x_{Gi} = \begin{bmatrix} v_{i,d} & v_{i,q} & i_{ti,d} & i_{ti,q} \end{bmatrix}^T,$$
$$u_{Gi} = \begin{bmatrix} v_{ti,d} & v_{ti,q} & i_{oi,d} & i_{oi,q} \end{bmatrix}^T. \quad (10)$$

The process and measurement noises are $w_{Gi}$ and $v_{Gi}$.

### B. Lines

The dynamical equations in *abc*-frame of line circuits are

$$\frac{d}{dt} i_{ij,abc} = -\frac{R_{ij}}{L_{ij}} i_{ij,abc} + \frac{1}{L_{ij}}(v_{i,abc} - v_{j,abc}). \quad (11)$$

Transferring to the same *dq0*-frame, it becomes

$$\frac{d}{dt} i_{ij,dq} + j\omega i_{ij,dq} = -\frac{R_{ij}}{L_{ij}} i_{ij,dq} + \frac{1}{L_{ij}}(v_{i,dq} - v_{j,dq}). \quad (12)$$

Similarly, their relation in steady state condition is

$$v_{i,dq} - v_{j,dq} = R_{ij} i_{ij,dq} + j\omega L_{ij} i_{ij,dq}. \quad (13)$$

From the above equations, the continuous-time state space model of lines of microgrids can be determined as:

$$\begin{cases} \dot{x}_{Lij}(t) = A_{Lij} x_{Lij}(t) + B_{Lij} u_{Lij}(t) + w_{Lij}(t) \\ \dot{x}_{Lik}(t) = A_{Lik} x_{Lik}(t) + B_{Lik} u_{Lik}(t) + w_{Lik}(t) \\ \ldots (i,j,k = 1\ldots n; i \neq j \neq k \ldots) \\ y_L(t) = C_L x_L(t) + v_L(t) \end{cases} \quad (14)$$

Where

$$A_{Lij} = \begin{bmatrix} -R_{ij}/L_{ij} & \omega \\ \omega & -R_{ij}/L_{ij} \end{bmatrix}, B_{Lij} = \begin{bmatrix} 1/L_{ij} & 0 \\ 0 & 1/L_{ij} \end{bmatrix}, \quad (15)$$

*n* is the number of buses in microgrids.

The state and input vectors are respectively as follows:

$$x_{Lij} = \begin{bmatrix} i_{ij,d} & i_{ij,q} \end{bmatrix}, u_{Lij} = \begin{bmatrix} v_{ij,d} & v_{ij,q} \end{bmatrix}. \quad (16)$$

Where, $v_{ij,dq} = v_{i,dq} - v_{j,dq}$. Similarly, $w_{Lij}$, $w_{Ljk}$ and $v_L$ are process and measurement noises, respectively.

### C. The Power Flow Equations

Unlike the polar form, the power flow equations do not contain orthogonal functions e.g. sin or cos as follows:

$$P_{oi} = \frac{3}{2}(v_{i,d} i_{oi,d} - v_{i,q} i_{oi,q})$$
$$Q_{oi} = \frac{3}{2}(v_{i,d} i_{oi,q} + v_{i,q} i_{oi,d}) \quad (17)$$

## III. DECENTRALIZED DYNAMIC STATE ESTIMATIONS

Based on the previously derived models, this Section will detail the proposed decentralized local estimators and global estimators.

### A. Kalman filter algorithm

Since the process and measurement noises are assumed to be zero-mean Gaussian distribution with covariance Q and R, the Kalman filter is a great algorithm to remove them. In various previous research, by dealing with the measurement value in the polar form, the Jacobian matrices should be computed to linearize the state-space model [1]. Recently, other advanced algorithms such as unscented Kalman [16] or particle filters [17] were implemented to overcome the problem of nonlinearity of the state-space model.

Assumed that the grids parameters are unchanged, the dynamic state-space model derived in (7) and (14) are linear and time-invariant (LTI). Hence, the standard Kalman filter can be easily applied.

Considering the deterministic discrete LTI state-space model

$$\begin{cases} x_{k+1} = A_d x_k + B_d u_k + w_k \\ y_k = x_k + v_k \end{cases}, \quad (17)$$

where *k* denotes the time instant $k^{th}$. At the sampling time $T_s$, the state *x* and input matrices can be derived based on the corresponding matrices $A_c$ and $B_c$ in continuous-domain. When $T_s$ is sufficiently small, the formulas in (18) can be used to approximate the discrete matrices; (19) is for the general cases assumed that $A_c$ is invertible.

$$A_d = I + T_s A_c; B_d = T_s B_c. \quad (18)$$

$$A_d = e^{T_s A_c}; B_d = (e^{T_s A_c} - I) B_c A_c^{-1}. \quad (19)$$

The following steps can be performed recursively to achieve the system state estimation.

*1) Predicting the states based on previous values:*
- Predict priori estimated states:

$$\hat{x}_{k|k-1} = A\hat{x}_{k-1|k-1} + Bu_k \quad (20)$$

- Predict priori error covariance:

$$P_{k|k-1} = AP_{k-1|k-1}A^T + Q \quad (21)$$

## 2) Updating to the estimated states::
- Calculate innovation:
$$r_{k|k-1} = \tilde{x}_k - \hat{x}_{k|k-1} \quad (22)$$

- Calculate innovation covariance:
$$S_k = R + P_{k|k-1} \quad (23)$$

- Kalman filter gain:
$$K_k = P_{k|k-1} S_k^{-1} \quad (24)$$

- Update estimated states:
$$\hat{x}_{k|k} = \hat{x}_{k|k-1} + K_k r_{k|k-1} \quad (25)$$

## 3) Checking errors, then return to step 1
- The post-fit residual:
$$r_{k|k} = \tilde{x}_k - x_{k|k} \quad (26)$$

- Update estimated covariance:
$$P_{k|k} = (I - K_k) P_{k|k-1} (I - K_k)^T + K_k R_k K_k^T \quad (27)$$

### B. Local Estimator

In the area of DGU, the distances between measurement units and the field controller that hosts the local estimator are not normally far. Therefore, the communication delay can be neglected. In addition, the state vector size here is just four as expressed in (10). Thus, the local estimation can have a sampling rate that is as high as possible with respect to the sensors.

Considering the discrete form of the state-space model described in (7) as the below equation (28), where the corresponding matrices were discretized.

$$\begin{bmatrix} v_{i,d} \\ v_{i,q} \\ i_{ti,d} \\ i_{ti,q} \end{bmatrix}_{k+1} = A_{Gdi} \begin{bmatrix} v_{i,d} \\ v_{i,q} \\ i_{ti,d} \\ i_{ti,q} \end{bmatrix}_k + B_{Gdi} \begin{bmatrix} v_{ti,d} \\ v_{ti,q} \\ i_{oi,d} \\ i_{oi,q} \end{bmatrix}_k \quad (28)$$

Notably, the inputs $v_{ti,dq}$ and $i_{oi,dq}$ are also measured or estimated, the accompanying noises of these measurements can degrade the performance of the Kalman filter. To eliminate the input noise effects, the following steps can be taken [18].

Assumed that the inputs are also associated with zero mean Gaussian noises as follows

$$\tilde{u}_k = u_k + m_k, \quad (29)$$

where $M$ is covariance of input noises. The discrete state-space model now can be described as:

TABLE I. ELECTRICAL PARAMETER OF MICROGRID

| Symbols | Values | Symbols | Values |
|---|---|---|---|
| $R_{t1}$ | 1.1 mΩ | $L_{t1}$ | 90 μH |
| $R_{t2}$ | 1.3 mΩ | $L_{t2}$ | 100 μH |
| $R_{t3}$ | 0.9 mΩ | $L_{t3}$ | 110 μH |
| $R_{12}$ | 1.1 Ω | $L_{12}$ | 0.52 mH |
| $R_{13}$ | 0.9 Ω | $L_{13}$ | 0.44 mH |
| $R_{23}$ | 1.3 Ω | $L_{23}$ | 0.67 mH |
| $V_{ll}$ | 13.8 kV | $C_{t1}$ | 50 μF |
| $f_n$ | 60 Hz | $C_{t2}$ | 55 μF |
| $T_s$ | 0.0001 | $C_{t3}$ | 60 μF |

$$x_k = A_d x_{k-1} + B_d u_k + B_d m_k + w_k \quad (30)$$

The process noise now became

$$\tilde{w}_k = B_d m_k + w_k \quad (31)$$

Since the input matrix in (9) is constant, then $B_d m_k$ is Gaussian. As a result, $\tilde{w}_k$ is also Gaussian. The mean and covariance of $\tilde{w}_k$ can be determined as:

$$\mathrm{E}(\tilde{w}_k) = B_d \mathrm{E}(m_k) + \mathrm{E}(w_k) = 0 \quad (32)$$

$$\begin{aligned} \mathrm{Cov}(\tilde{w}_k) &= \mathrm{Cov}(B_d m_k) + \mathrm{Cov}(w_k) \\ \mathrm{Cov}(\tilde{w}_k) &= \tilde{Q} = B_d M B_d^T + Q \end{aligned} \quad (33)$$

Now, the classical Kalman filter process as described in Subsection III.A can be applied to this new process noise covariance $\tilde{Q}_k$.

### C. Global Estimator

The state vectors achieved from local estimators and independent sensors are sent to the control center to estimate the states of the entire microgrid. The state-space model in (14) is employed for this estimation provided that the measurement matrix is observable.

Note that in this model, the bus voltages should be known and considered as the input and the line currents are the system states. Since these relationships are independent, the estimation process can be performed in a nearby field controller to lessen the computational burden in the global estimation.

## IV. SIMULATION RESULTS

The three-bus microgrid is simulated in Matlab/Simulink with secondary and primary controls [19] for DGU. The electrical parameters of the system are given in Table I. The white noise with a certain covariance is added to the measurement vector. The load change event (disturbance) is generated at bus 1 at 2s to verify the dynamic response.

The local estimation is processed with the true value being sampled at 10 kHz. The comparison between the true, the noisy, and the estimated values of $v_{d1}$, $v_{q1}$, $i_{d1}$, and $i_{q1}$ are shown in Figs. 3, 4, 5, and 7, respectively. The zoomed picture of Figs. 5 and 7 are shown in Figs. 6 and 8, respectively.

The global estimation is processed with the true value being sampled at 100 Hz. The comparison between the true, the noisy, and the estimated values of $i_{d12}$ and $i_{q12}$ are shown at Figs. 9 and 10, respectively.

As can be seen in Figs. 3-10, the estimated value is around the true value with small errors. When the load change event occurred in 3s, the estimators can still track the true value of the system states.

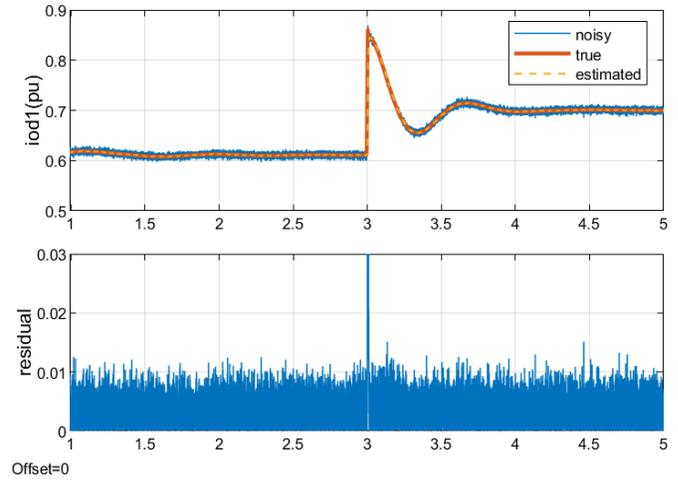

Figure 5. Comparison of estimated, noisy and true value of $i_{d1}$

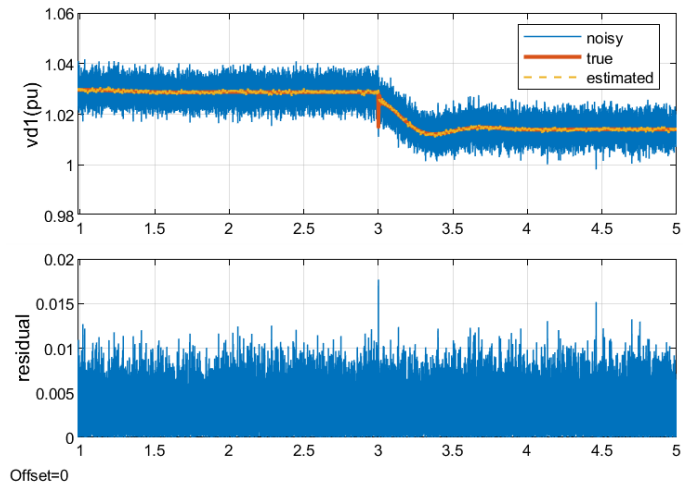

Figure 3. Comparison of estimated, noisy and true value of $v_{d1}$

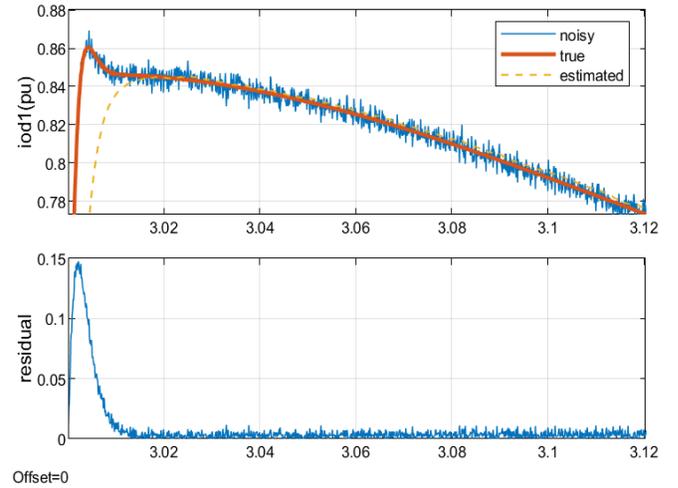

Figure 6. Zoomed comparison of estimated, noisy and true value of $i_{d1}$

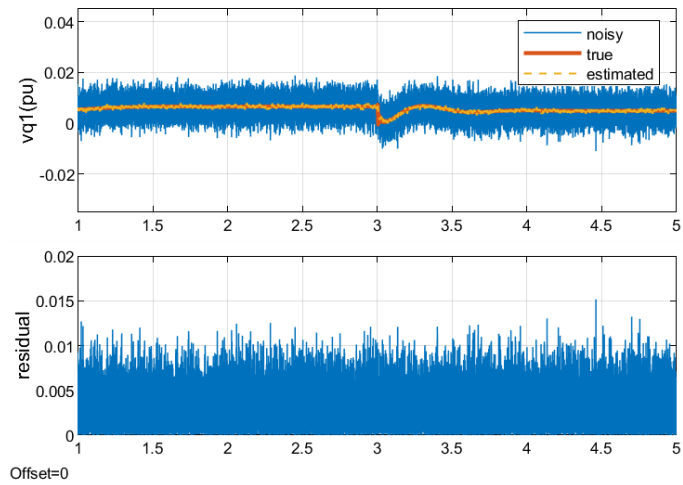

Figure 4. Comparison of estimated, noisy and true value of $v_{q1}$

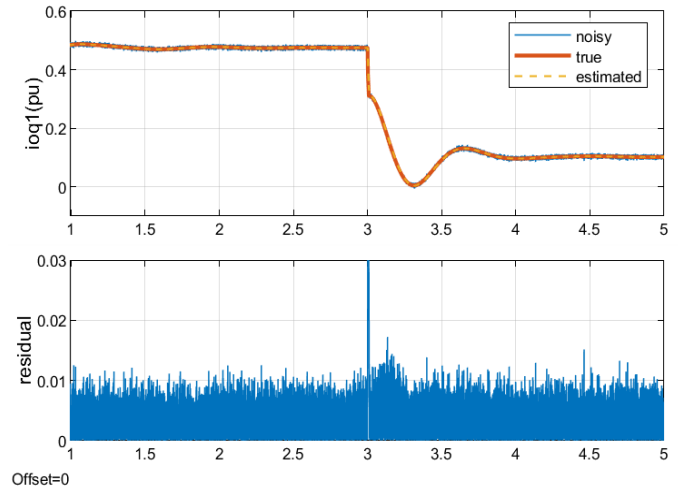

Figure 7. Comparison of estimated, noisy and true value of $i_{q1}$

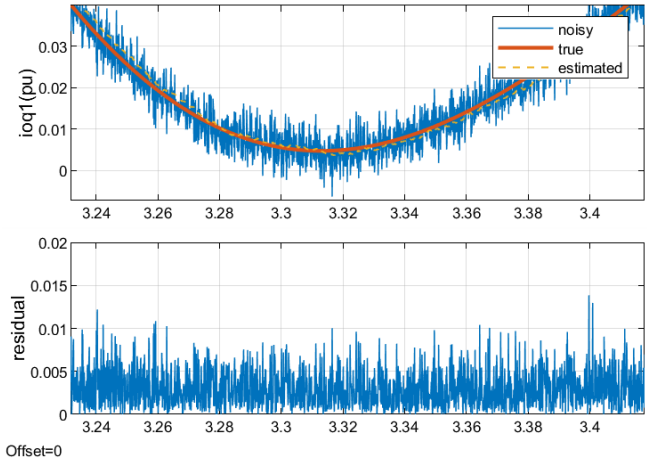

Figure 8. Zoomed comparison of estimated, noisy and true value of $i_{q1}$

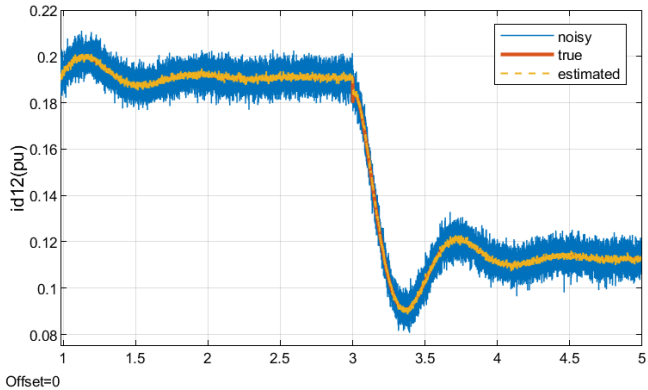

Figure 9. Comparison of estimated, noisy and true value of $i_{d12}$

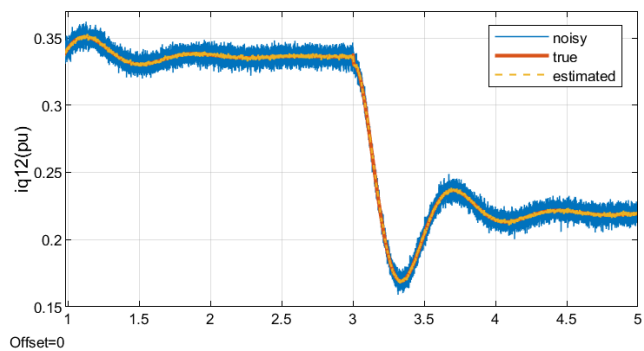

Figure 10. Comparison of estimated, noisy and true value of $i_{q12}$

## V. CONCLUSION

A simple decentralized dynamic state estimation scheme based on the same dq0-rotating reference frame in microgrids is proposed and analyzed. The advantage of this scheme over the previous research in the same category is linear formulation with dynamic state-space models. In addition, by revising the process noised, the noisy inputs do not affect the state estimation process. The classical Kalman filter is employed, thus the computational time and effort are reduced significantly. Simulation results are provided for verification. In this work, µPMU were employed in the proposed method; however, future MVDC ship systems will not have the phasor characteristic that µPMU can be utilized for. Our future approach to the future classes of state estimation for MVDC ship power systems will be focusing on state estimation on other types of advanced sensor network.